\magnification=1095
\raggedbottom
\overfullrule=0pt
\headline={\ifnum\pageno>1 \hfil -- \folio\ -- \hfil \else\hfil\fi}
\footline={}
\vsize=9truein
\font\smc cmcsc10 at 12 truept
 at 16 truept
\font\lbf cmbx12 at 16 truept
 at 10 truept
 at 10 truept
\def\ref#1{{\par\noindent \hangindent=3em\hangafter=1 #1\par}}  
\def\ditto{\vrule width1.2cm height3pt depth-2.5pt}
\def\largeskip{\vskip 0.25 truein}                              

\def\ft#1{$^{\;\rm #1}$}
%

\def\Msol{\ifmmode{\rm M}_{\mathord\odot}\else M$_{\mathord\odot}$\fi}
\def\deg{$^{\circ}$}

\def\kms{km~s$^{-1}$}
\def\m#1{$^{-#1}$}
\def\tten#1{$\times 10^{#1}$} 
\def\ten#1{$10^{#1}$} 

\def\ls{\lower 2pt \hbox{$\;\scriptscriptstyle \buildrel<\over\sim\;$}} 
\def\gs{\lower 2pt \hbox{$\;\scriptscriptstyle \buildrel>\over\sim\;$}} 

\def\a{$\alpha$}                                          
\def\b{$\beta$}
\def\l{$\lambda$}
\def\tl{$\lambda\lambda$}
\def\i{{\smc~i}}
\def\ii{{\smc~ii}}
\def\iii{{\smc~iii}}

\newcount\Q
\Q=0

\newcount\F
\F=0
\advance\F by 1 \newcount\Fprof \Fprof=\F 
\advance\F by 1 \newcount\Frvel \Frvel=\F 
\advance\F by 1 \newcount\Fpec \Fpec=\F 
\advance\F by 1 \newcount\Ffit  \Ffit=\F  
\advance\F by 1 \newcount\FGask \FGask=\F 

\input psfig
\font\bigsf=cmssbx10 scaled 1400
\hbox{
\psfig{figure=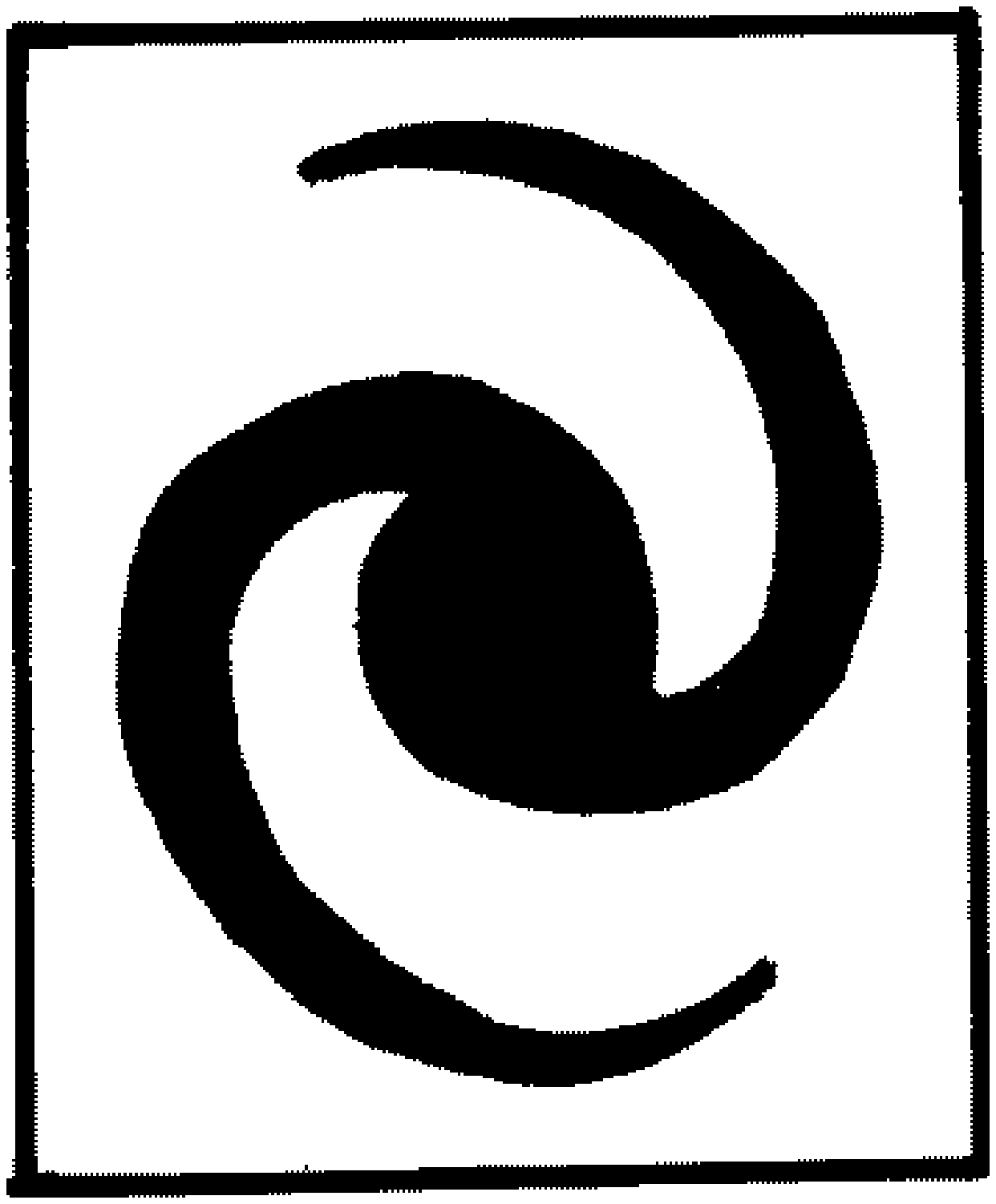,height=1in,width=1in}
\hskip 0.5truein 
\vbox to 1. truein {\bigsf \vfill
\noindent UNIVERSITY OF CALIFORNIA AT BERKELEY 
\medskip \noindent ASTRONOMY DEPARTMENT
\vfill}
}

\largeskip

\centerline{\lbf REJECTION OF THE BINARY BROAD-LINE REGION}
\medskip
\centerline{\lbf INTERPRETATION OF DOUBLE-PEAKED EMISSION}
\medskip
\centerline{\lbf LINES IN THREE ACTIVE GALACTIC NUCLEI}

\largeskip
\centerline{\smc Michael Eracleous\ft{1,2}, Jules P. Halpern\ft{2,3}, Andrea
            M. Gilbert,}
\centerline{\smc Jeffrey A. Newman, and Alexei V. Filippenko}
\bigskip
\centerline {Department of Astronomy, University of California, Berkeley, CA 94720}
\bigskip
\centerline{To appear in the {\it Astrophysical Journal}}

\footnote{}{\parindent=0pt
\item{\ft{1}} Hubble Fellow, e-mail: {\tt mce@beast.berkeley.edu}.
\smallskip \item{\ft{2}}
Visiting Astronomer, Kitt Peak National Observatory, which is operated by AURA,
Inc., under a cooperative agreement with the National Science Foundation.
\smallskip \item{\ft{3}} 
Permanent address: Columbia Astrophysics Laboratory,
Columbia University, 538 West 120th Street, New York, NY 10027.
}

\largeskip

\centerline {ABSTRACT}
\bigskip
\noindent
It has been suggested that the peculiar double-peaked Balmer lines of certain
broad-line radio galaxies come from individual broad-line regions associated
with the black holes of a supermassive binary. We continue to search for
evidence of the radial velocity variations characteristic of a double-lined
spectroscopic binary that are required in such a model. After spectroscopic
monitoring of three suitable candidates (Arp~102B, 3C~390.3, and 3C~332)
spanning two decades, we find no such long-term systematic changes in radial
velocity. A trend noticed by Gaskell (1996$a$) in one of the Balmer-line peaks
of 3C~390.3 before 1988 did not continue after that year, invalidating his
inferred orbital period and mass. Instead, we find lower limits on the plausible
orbital periods that would require the assumed supermassive binaries in all
three objects to have total masses in excess of $10^{10}$~\Msol. In the case
of 3C~390.3 the total binary mass must exceed \ten{11}~\Msol\ to satisfy
additional observational constraints on the inclination angle.  We argue that
such large binary black  hole masses are difficult to reconcile with other
observations and with theory.  In addition, there are peculiar properties of the
line profiles and flux ratios in these objects that are not explained by
ordinary broad-line region cloud models. We therefore doubt that the
double-peaked line profiles of Arp~102B, 3C~390.3, and 3C~332 arise in a pair of
broad-line regions. Rather, they are much more likely to be  intimately
associated with a single black hole. The recent discoveries of transient but
otherwise similar double-peaked emission lines in nearby AGNs bolster the view
that double-peaked emission lines are commonly produced by a single compact
source.

\medskip\noindent
{\it Subject headings:} galaxies: active -- galaxies: individual (Arp~102B,
3C~390.3, 3C 332) -- lines: profiles

\vfill
\eject

\centerline {\smc 1. introduction}
\medskip
The possibility that the nuclei of radio galaxies may harbor supermassive binary
black holes was suggested by Begelman, Blandford, \& Rees (1980) with the
purpose of explaining the observed precession of radio jets in these objects. 
In the original scenario, the supermassive binary would form as a result of the
merger of two parent galaxies, each with its own nuclear black hole. Because
radio galaxies are preferentially ellipticals, which are commonly thought to be
the products of mergers, this scenario offers an appealing explanation for the
proposed association of binary black holes with the nuclei of radio galaxies.
The tidal perturbation of one black hole on the jet emanating from the other was
the proposed origin of the jet precession.  A similar idea was invoked by Wilson
\& Colbert (1995) who proposed that the formation of radio jets is associated
with the merger of a supermassive binary. The exploration of the formation and
evolution of a supermassive binary by Begelman et al. (1980) showed that for
conditions typical of the nucleus of an elliptical galaxy, the supermassive
binary spends most of its life (which could range between $10^8$  and
$10^{10}$~yr for black hole masses on the order of $10^8\,\Msol$) at a
separation of the order of $0.01-0.1$~pc.

In addition to perturbing the radio jet, a second nuclear black hole may affect
the dynamics of the broad-line emitting gas and hence leave an imprint on the
profiles of the emission lines, as pointed out by Begelman et al.  (1980) and
Gaskell (1983). In particular, broad emission lines that originate in gas that
is closely bound to either of the two black holes will be displaced in velocity
from the rest frame of the host galaxy (as defined by either its narrow emission
lines or its stellar absorption features) because of the orbital motion of the
black holes about their center of mass.  Gaskell (1983, 1988, 1996$a$)
identified several broad-line radio galaxies as candidate hosts of supermassive
binaries on the basis of either single or double displaced peaks in their broad
emission-line profiles, and proposed that the binary model be applied generally
to displaced emission-line peaks in active galactic nuclei (hereafter AGNs). 
Going one step further, Gaskell (1996$b$) and Gaskell \& Snedden (1997)
suggested that a large fraction of radio-loud as well as radio quiet AGNs may
harbor binary  black holes, objects with clearly double-peaked emission
lines  being the most obvious examples. The binary black hole interpretation was
also discussed in favorable terms by Stockton \& Farnham (1991) for the
double-peaked Balmer lines of the quasar OX~169, in part because they discovered
that its host galaxy appears to be an advanced merger. Additional radio-loud
AGNs with double-peaked emission lines that might be treated as candidate
supermassive binaries can be found in Eracleous \& Halpern (1994). Those
authors, however, favored an origin in a single accretion disk for the majority
of the double-peaked line profiles that they found.

The most definitive observational evidence that a binary broad-line region is
responsible for a double-peaked emission line would be the opposite drift of its
two peaks as a result of the orbital motion of the binary black hole. Whether
such a drift would be observable or not depends on the orbital period of the
binary, which could range from decades to centuries.  Halpern \& Filippenko
(1988, 1992) searched for evidence of orbital motion in two double-peaked
emitters, Arp~102B and 3C~332, without success. The observed peak velocities and
lower limits to their periods translate into lower limits on the masses of the
binary black holes that could be responsible for binary broad-line regions in
these objects.  For Arp~102B and 3C~332, the derived lower limits were
4\tten{9}~\Msol\ and 2\tten{10}~\Msol, respectively.  It was noted that
if continued monitoring of the double-peaked emission lines in these two objects
failed to detect orbital motion, then the mass needed under the binary 
broad-line region interpretation would be constrained to even larger values,
ultimately rendering the binary broad-line region hypothesis untenable.  In
contrast to these null results, a positive result on a third broad-line radio
galaxy was recently claimed by Gaskell (1996$a$), who found a systematic
velocity variation in one of the two peaks of the H\b\ line of 3C~390.3. This
was interpreted as evidence of orbital motion with a most likely period of
300~yr and a binary mass of 6.6\tten{9}~\Msol .

Our continuing observational test of the binary broad-line region hypothesis for
all three of these objects (Arp~102B, 3C~390.3, and 3C~332) is the subject of
this paper. We use all the available data, increasing the time base on each of
the three objects by  at least 6 years over previously published results, thus
achieving the more restrictive constraints that were anticipated by Halpern \&
Filippenko (1988). We discuss our (null) results in the context of other
observations and theories that might be relevant to the binary black hole
interpretation of double-peaked emission lines.

\bigskip
\centerline {\smc 2. the observational test}
\medskip

In the context of the binary black hole hypothesis, and by analogy with stellar
double-lined spectroscopic binaries, we can suppose that the two peaks in
double-peaked emission lines originate in ionized gas in the vicinity of each of
the two black holes.  We can test this supposition by searching for orbital
motion of the black holes and their associated emission-line regions. Even if
orbital motion is not definitely seen, a very important constraint that can be
derived from the variation of the velocities of the emission-line peaks, or the
lack thereof, is a lower limit on the total mass of the hypothesized binary. The
two black holes are assumed to follow circular orbits around their common center
of mass.  Orbits are likely to be circular because dynamical friction between
two parent galaxies during the merger ensures that the initial eccentricity of
the resulting black hole binary is small (Polnarev \& Rees 1994). An
eccentricity that is small initially will not grow, and it is likely to decrease
as the binary hardens (Quinlan 1996).  Moreover, if the eccentricity is
initially large the decay of the orbit by gravitational radiation will be
accelerated, with the consequence that binaries with eccentric orbits will be
much less common than ones with circular orbits (Quinlan 1996, and references
therein). Any stars that happen to be in the vicinity of the supermassive binary
will have a negligible effect on the potential governing the
orbit of the binary. Recent estimates of 
the stellar mass densities in the cores of nearby, late-type galaxies ($5\times
10^4 - 10^6$~\Msol~pc\m3; Lauer et al. 1991, 1992, 1995) imply that the mass 
of stars within the volume occupied by the binary would not exceed
$10^6$~\Msol. This is several orders of magnitude smaller than the black hole
masses that we estimate in the following sections.

We assume that the two black holes have masses and circular orbital velocities
$M_1, v_1$ and $M_2, v_2$, respectively, with a mass ratio
$q=M_1/M_2 \geq 1$ by definition. It follows that the orbital velocities are
related by $v_1/v_2=1/q$. We also define the total mass of the binary as
$M=M_1+M_2$, which is related to the orbital period
$P$ and the velocities as
\newcount\QmassA \advance\Q by 1 \QmassA=\Q
\newcount\QmassB \advance\Q by 1 \QmassB=\Q
$$
\eqalignno{
M = & \; {1\over 2\pi G}\; \left(1+q\right)^3\; P \; v_1^3 & (\the\QmassA) \cr
{\rm or} \quad M = & \; {1\over 2\pi G}\; \left({1+q\over q}\right)^3\; P \;
v_2^3. & (\the\QmassB) \cr}
$$
Unfortunately, the observer cannot measure the true orbital velocities, but
instead only their projections along the line of sight, $v_1\sin i$ and $v_2\sin
i$ (where $i$ is the angle between the line of sight and the normal to the
plane of the orbit). Assuming that the mass ratio can be measured from the line
profile and that lower limits on the period and the projected orbital velocities
can be obtained, equations (\the\QmassA) and (\the\QmassB) yield lower limits on
the total mass of the binary as follows:
\newcount\QmlimA \advance\Q by 1 \QmlimA=\Q
\newcount\QmlimB \advance\Q by 1 \QmlimB=\Q
$$\eqalignno{
M > & \;
4.7 \times 10^8 \left(1+q\right)^3\; \left({P\over {\rm 100\; yr}}\right)
\;
\left({v_1\; \sin i\over 5000\;\; {\rm km\;\; s^{-1}}} \right)^3
\;\; \Msol & (\the\QmlimA) \cr
{\rm or} \quad M > & \;
4.7 \times 10^8 \left({1+q\over q}\right)^3\;
\left({P\over {\rm 100\; yr}}\right) \; \left({v_2\; \sin i \over 5000\;\; {\rm
km\;\; s^{-1}}} \right)^3\;\; \Msol. & (\the\QmlimB)\cr
}
$$
Even if the mass ratio cannot be measured, which might be the case if one of the
peaks in the line profile is weak, absent, or contaminated by unrelated narrow
emission-lines, a somewhat smaller (more conservative) lower limit on the total
mass can still be derived because $q \geq 1$ by definition, so $(1+q)^3 \geq 8$
and $[(1+q)/q]^3
\geq 1$.

We apply the standard analysis of a double-line spectroscopic binary
to measurements of the double-peaked H\a\ lines to test the binary
broad-line region hypothesis. Examples of the line profiles of the
three objects on which this study focuses are presented in
Figure~\the\Fprof. H\a\ is the line of choice because it is the
strongest double-peaked line, and because both of its peaks are often
uncontaminated by narrow lines and hence are readily measurable.  We
measure the locations of both H\a\ peaks wherever possible to maximize
the sensitivity of the test. The H\b\ lines were also used, whenever
available, to measure the velocity of the blue peak.  Under the
double-lined spectroscopic binary hypothesis, the peak with the larger
velocity displacement relative to the velocity of the host galaxy is
associated with the less massive black hole. The ratio of the
velocities of the two peaks gives the mass ratio of the binary and is
required to remain constant as the binary components revolve. The
period and amplitude of the radial velocity curves can be fitted,
although in practice only lower limits to both of these quantities
have thus far been obtained. Such constraints are nevertheless useful
because they yield lower limits to the hypothesized mass.
\topinsert
\centerline{\hbox to 6.truein{ \hskip -0.5 truein
\psfig{figure=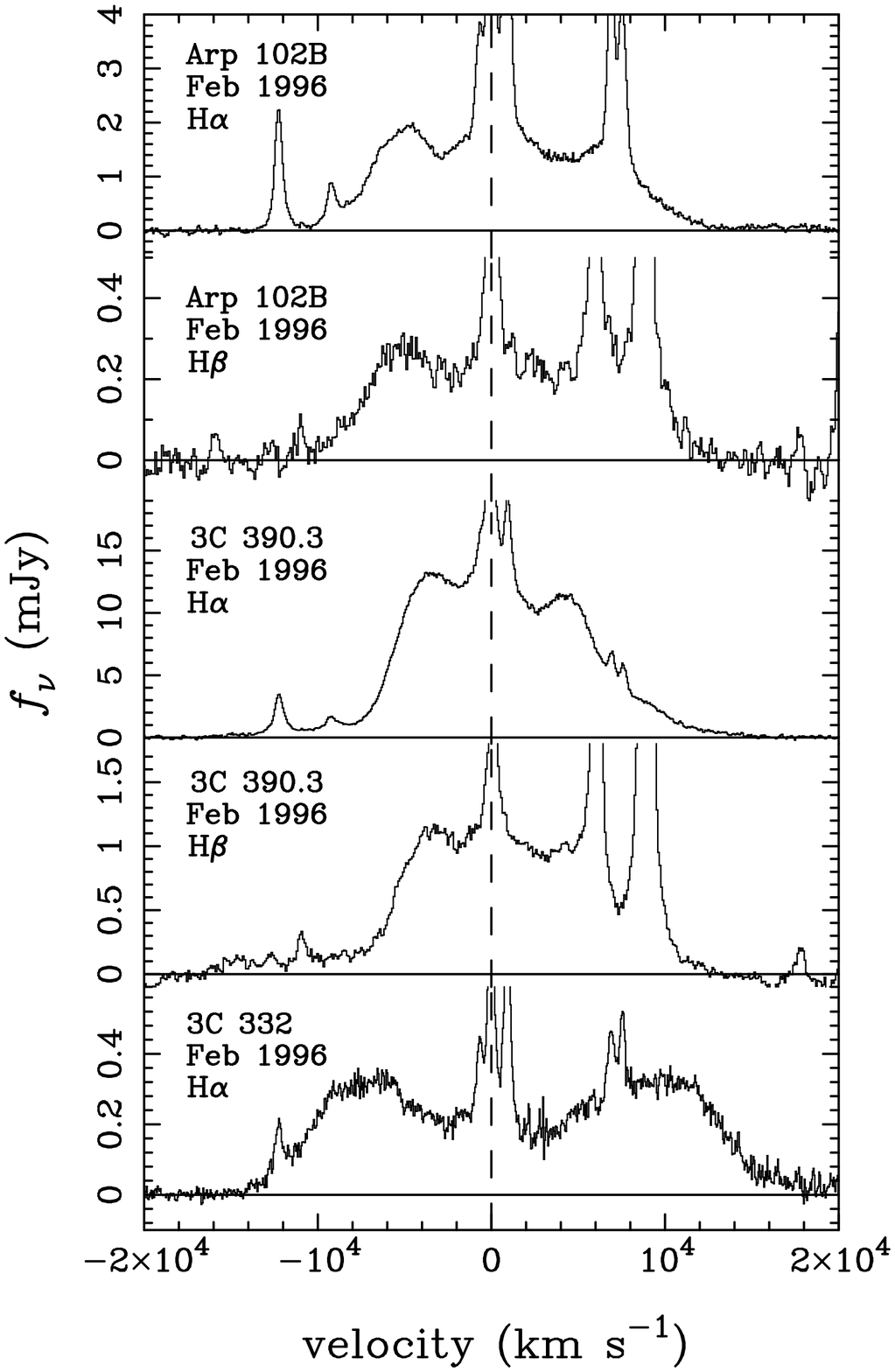,height=4in,rheight=3.8in,rwidth=3in}
\hfill
\vbox to 4 truein {\hsize=3. truein
\noindent {\smc Figure \the\Fprof -- } 
Examples of the H\a\ and H\b\ profiles of the three target objects: Arp~102B,
3C~390.3, and 3C~332. The particular spectra shown here were obtained with the
Kitt Peak National Observatory's 2.1~m telescope in 1996 February. The H\b\ line
of 3C~332 is not included in this figure because it is very weak and cannot
provide accurate measurements of the peak velocities. \vfill
}}}
\endinsert

\bigskip
\centerline {\smc 3. The Data: Spectra, Line Profiles, and Measurements}
\medskip

The observational data consist of optical spectra of the three target objects
covering their H\a\ and/or H\b\ lines.  We have been accumulating spectra of
these objects for about 10 years at a rate of about once per year initially and
up to several times per year more recently. Our observations were carried out at
Palomar Observatory, Lick Observatory, the Michigan-Dartmouth-MIT Observatory,
and Kitt Peak National Observatory.  A compilation of many of these spectra is
shown in Eracleous (1997). Spectra taken at Lick Observatory by various
observers in the 1970s and 1980s were used to supplement our own collection. In
particular, we use spectra of 3C~332 obtained by M. M. Phillips and S. A. Grandi
in 1974 and 1976, and several spectra of 3C~390.3 taken by D. E. Osterbrock
between 1974 and 1988. The early Lick observations of 3C~332 were originally
reported by Grandi \& Osterbrock (1978) and Grandi \& Phillips (1979), while 
the early Lick spectra of 3C~390.3 were presented by Zheng et al. (1995).
Measurements were also made from additional published spectra, namely the 1982 spectrum of
Arp~102B presented by Stauffer, Schild, \& Keel (1983), and the 1981 and 1983
spectra of 3C~390.3 presented by Oke (1987).

In Figure~\the\Fprof\ we show examples of the H\a\ and H\b\ line profiles of the
three target objects. The H\b\ line of 3C~332 is not included in this figure
because it is generally very weak, and thus cannot be used for accurate
measurements of the peak locations. As Figure~\the\Fprof\ shows, the H\a\ lines
are considerably stronger than the H\b\ lines and hence the H\a\ profiles have a
much higher signal-to-noise ratio. It is true that the telluric A- and B-bands
often contaminate the H\a\ line; these absorption features can routinely be
corrected, however, to an accuracy that fully recovers the intrinsic H\a\ line
profile with the help of spectra of featureless standard stars (e.g.,
Wade \& Horne 1988).  H\a\ is thus the preferred line for measuring the
velocities of the twin peaks. The red peak of H\b\ coincides with the [O\iii]
\tl4959, 5007 doublet and hence is not measurable. However, the H\b\ line is
still useful since the location of its blue peak can be measured and tested for
consistency with H\a . We find that the profiles of the H\a\ and H\b\ lines are
almost identical, so measurements of the blue peak of H\b\ can supplement those
of H\a. In some of the older spectra of 3C~390.3 obtained at Lick Observatory,
the telluric B-band at 6867~\AA\ is not corrected well, making measurements of
the blue peak of H\a\ unreliable. In the case of Arp~102B the red peak of H\a\
cannot be measured because it coincides with the [S\ii]
\tl6717, 6731 doublet.

We  measure the velocities of the displaced peaks by fitting a symmetric,
bell-shaped function to the region of the profile around the peak. By
experimenting with Gaussian and parabolic fits we found that the two produce
almost identical results; we adopted the Gaussian fitting method. This method
locates an effective flux-weighted centroid of the profile, which does not
necessarily coincide with the highest point (mode), especially if the profile is
skewed. The formal uncertainty in the location of the peak is typically
100~\kms\ for profiles with high signal-to-noise ratio. The actual uncertainty
may, however, be dominated by systematic effects of fitting a skewed profile
with a symmetric function. We will return to this issue in our later discussion
of the measurements. Gaskell (1996$a$) employed Pogson's method for measuring
the velocities of the displaced peaks (Hoffmeister,
Richter, \& Wenzel 1985). Gaskell's implementation of this method (C. M.
Gaskell, private communication) involves finding the midpoint between the wings
(or sides) of the profile at several levels below the peak, and then locating
the ``center'' of the profile by averaging these midpoints (in a graphical
application of this technique the midpoints are averaged by fitting them 
with a straight line). This procedure avoids the peak of the profile because
it is often skewed. Any noise in the wings of the profile can also affect the
location of its peak by this method. We have compared the
results of our Gaussian fitting method with the results of Gaskell's version of
Pogson's method using several spectra of 3C~390.3 as examples. We find that
the two methods give results that are in good agreement with each other. In the
cases that we have tested, the mean discrepancy between the two methods is 
0.6~\AA, corresponding to 30~\kms, which is smaller than the typical
uncertainty with which either method can locate the velocity of a displaced
peak. This justifies combining our data with those of Gaskell (1996$a$) in 
a single set. In Figures~\the\Frvel$a$ and ~\the\Frvel$b$ we present
the individual measurements of the peak velocities of Arp~102b and 3C~390.3,
respectively. In the latter figure, the data  used by Gaskell (1996$a$), which
represent annual averages rather than individual measurements, are superposed
for comparison. In the case of 3C~332 we only use measurements made with the
Gaussian fitting method. The data are presented and discussed in the next
section. The spectra and tables of measured peak velocities are presented in
separate papers in which the variability of individual objects is studied in
much more detail. In particular, the Arp~102B data are given by Newman et
al. (1997), while the 3C~390.3 and 3C~332 data are given by Gilbert et al.
(1997).
\topinsert
\vskip -0.4 truein
\hbox to 6.5truein {
\psfig{figure=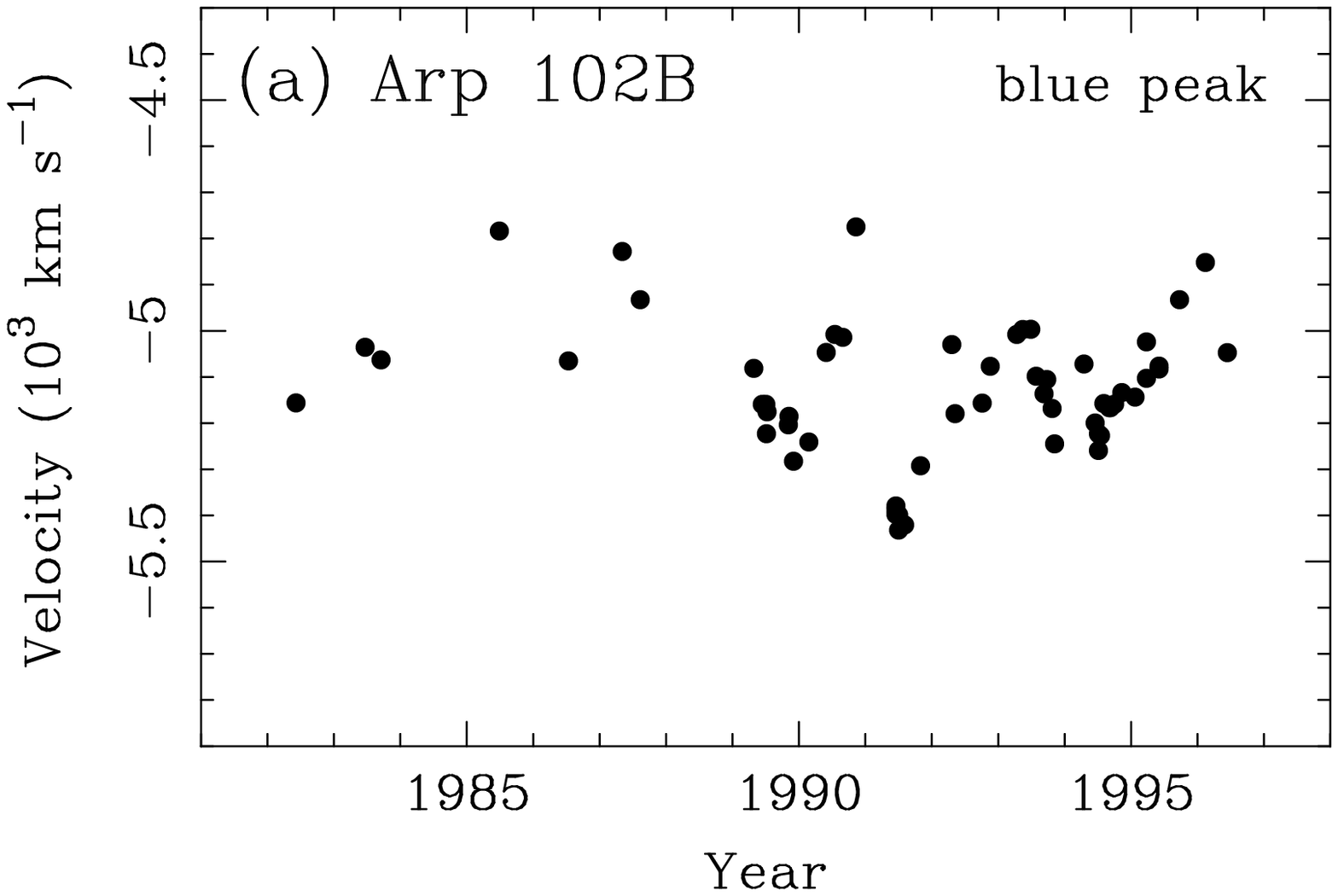,height=4in,rheight=3.8in}
\hfill
\psfig{figure=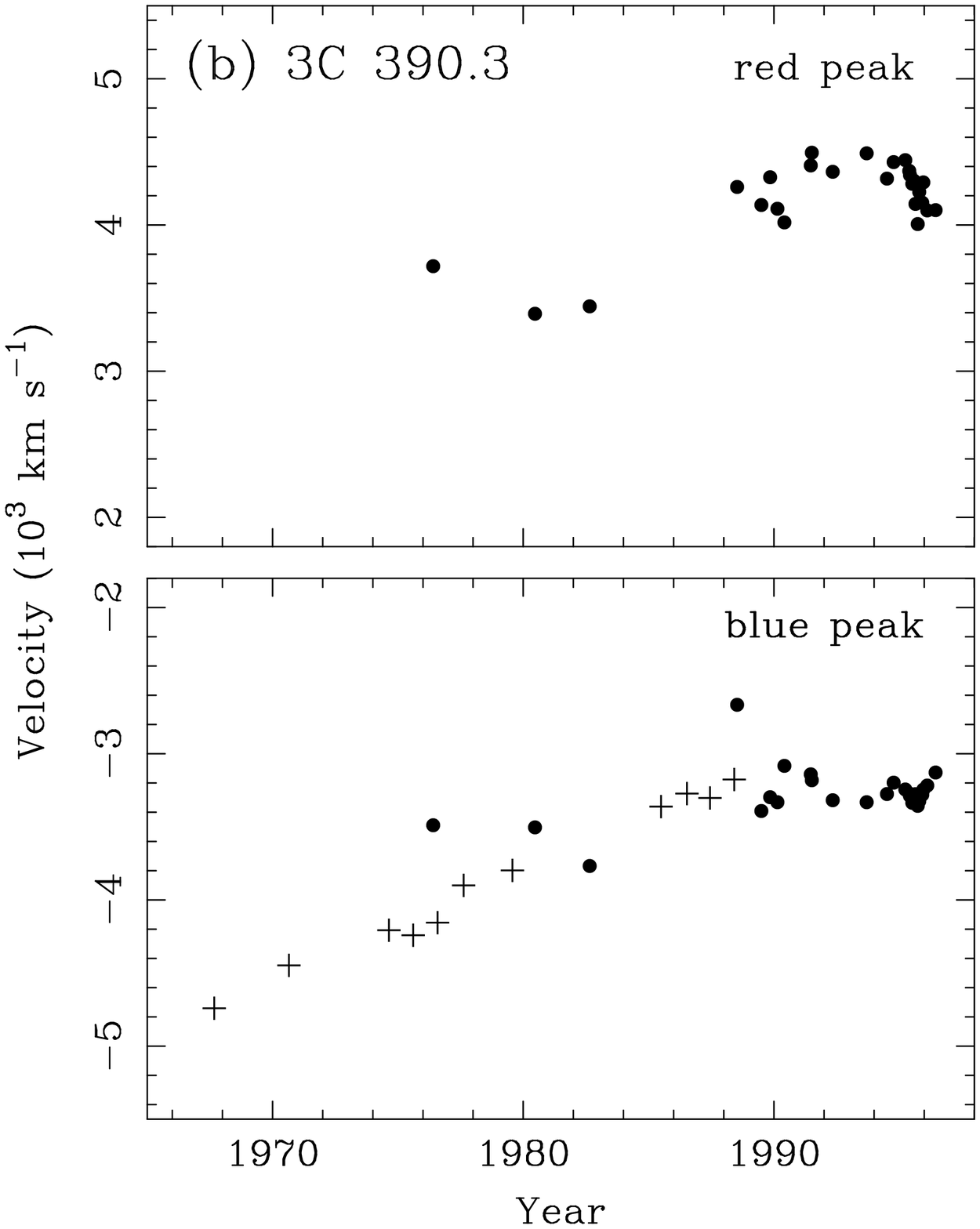,height=4in,rheight=3.8in}
}
\centerline{\vbox{\hsize=5.5truein
\noindent {\smc Figure \the\Frvel -- } 
Radial velocity curves of Arp~102B and 3C~390.3 based on measurements from our
spectra. Each point represents a separate measurement from an individual
H\a\ profile. Measurements made from the H\b\ profiles are omitted for clarity
Error bars are of the order of 100~\kms. The crosses in (b) show the
data used by Gaskell (1996$a$) and are included for comparison. They represent
annual averages rather than individual measurements.
}}
\medskip
\endinsert

The radial velocities in Figure~\the\Frvel\ show a significant scatter
on short time scales (a year or less). This is a result of intrinsic changes in
the detailed shape (skewness) of the profile rather than measurement
uncertainties. We demonstrate this by an example below. Such profile changes
trace the behavior of the line-emitting gas on time scales which are much
shorter than the orbital period of the supposed supermassive binary. They could
represent the reverberation of a variable ionizing continuum in the broad-line
region or redistribution in phase space of line-emitting gas close to one of the
two black holes on the local dynamical time scale.  Under these conditions we
are justified in computing annual velocity averages and using the annual
dispersion as a measure of the corresponding uncertainty.  This practice was
also adopted by Gaskell (1996$a$) who presented annual average velocities of the
blue H\b\ peak of 3C~390.3 between 1968 and 1989. By computing annual average
velocities, it is thus also possible to combine our data set with that of
Gaskell (1996$a$) and extend the temporal baseline spanned by the data on
3C~390.3. During the period 1989--1990, the line profile of 3C~390.3 displayed a
very sharp blue peak as shown in Figure~\the\Fpec. This is a somewhat extreme
but illustrative example of the type of intrinsic line profile changes that we
noted above. In Figure~\the\Fpec\ we overlay scaled versions of the 1988 and
1990 H\a\ profiles of 3C~390.3 for comparison with the 1989 profile. The very
shape of the 1989 profile warns against an interpretation in terms of two
otherwise ordinary broad-line regions, a point to which we will return in \S4.
However, in order to pursue the spectroscopic binary model of Gaskell (1996$a$)
to its logical conclusion, we apply the chosen measurement techniques to this
spectrum as well.
\topinsert
\vskip -0.6truein
\centerline{
\psfig{figure=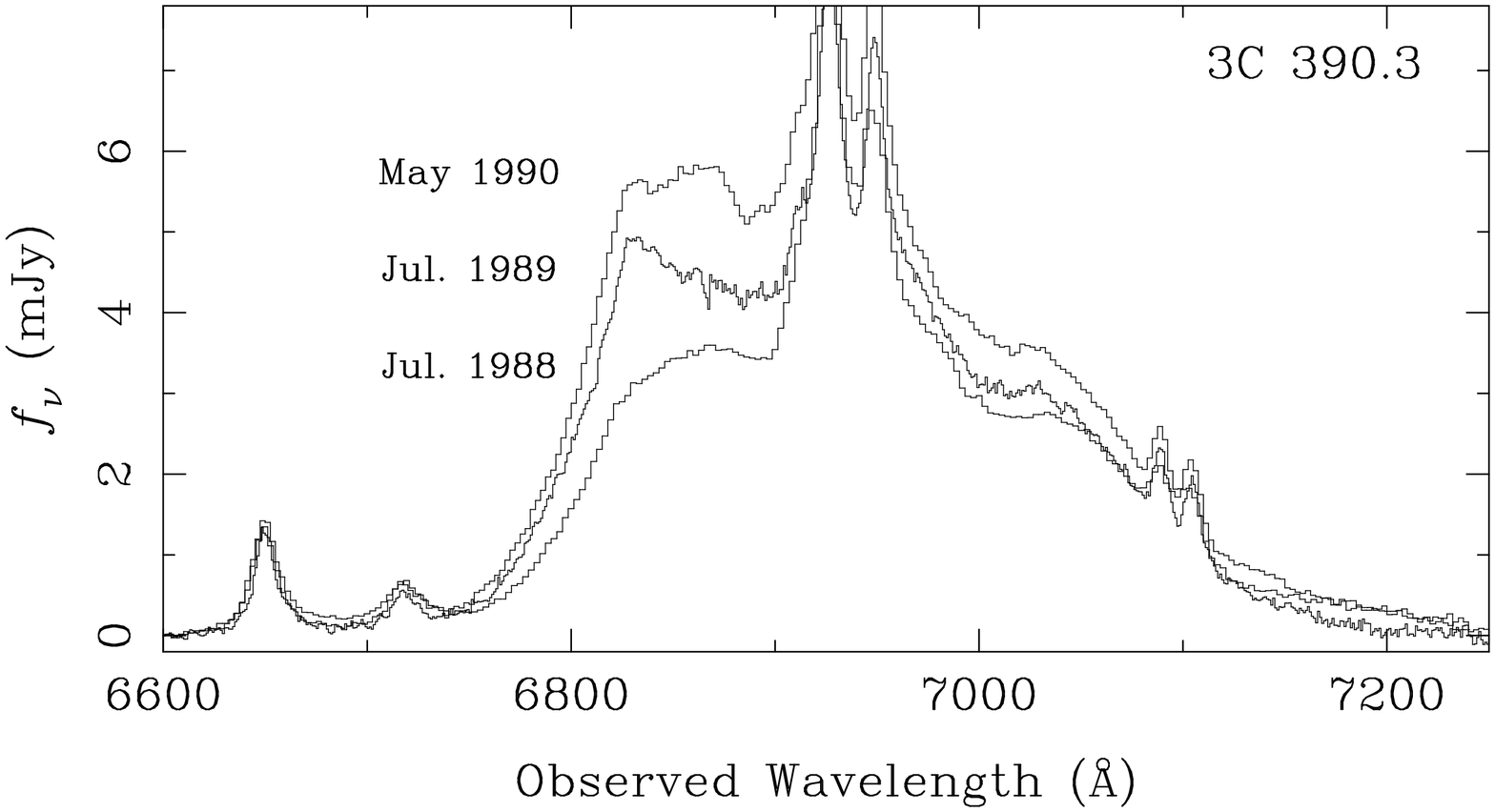,width=5in,rheight=3.3in,angle=-90}
}
\centerline{\vbox{\hsize=5.5truein \noindent {\smc Figure \the\Fpec\ -- }
H$\alpha$ spectrum of 3C~390.3 obtained in 1989 July, illustrating the very
sharp shelf or cusp at about 6830~\AA\ that dominates the structure of the
blueward displaced emission-line peak. For comparison we overlay the 1988 and
1990 H\a\ profiles, after scaling them to an  approximately equal flux in the
[O\i] \l6300 line.  In addition to demonstrating the 
dramatic change in the shape of the profile on a short time scale, this figure 
also illustrates how changes in the detailed shape of the peak can affect the 
measurement of its ``velocity.''
}}
\medskip
\endinsert

As we have noted above, the scatter in the measured velocity of a displaced
peak within a year is dominated by intrinsic changes in its shape. These
fluctuations have an unknown distribution, which is unlikely to be Gaussian;
hence the standard error in the mean, computed according to a Gaussian theory of
errors, would not reflect the true uncertainty. Thus, we have adopted the
root-mean-square (r.m.s.) dispersion, computed as 
$\sigma=(\langle v^2 \rangle - \langle v \rangle ^2)^{1/2}$, as a measure
of the uncertainty in the velocity of the hypothesized black hole, rather than
the error in the mean\ft{4}. 
\footnote{}{\parindent=0pt \item{\ft{4}}
The error bars of Gaskell (1996$a$) were computed as $\sigma_{\rm
n}=\sigma/\sqrt{n}$ (where $n$ is the number of measurements made within the
same year); they represent the error in the mean rather than the dispersion.
Because the dispersion in the annual samples of Gaskell's data is dominated by
measurement errors (C. M. Gaskell, private communication), we have retained 
these error bars without adjusting them to our convention.} 
In years when only one or two measurements are available, we assigned
the same error bar as in the year with the best-determined dispersion
(i.e., the largest number of measurements).  The error bars computed
as described above were used in our further analysis, including the
application of the $\chi^2$ test. Because the adopted error bars are
larger than the error in the mean (by a factor of 3 or less), the
resulting values of $\chi^2$ in the analysis presented below are
correspondingly lower.  As a result the lower limits on the binary
masses and periods that we ultimately derive are smaller (i.e. more
conservative) than what the error-in-the-mean error bars would
yield. The annually-averaged radial velocity curves were compared with
the predictions of the binary broad-line region model as we describe
in the next section, with the goal of deriving constraints on the
properties of the hypothesized binary black hole.

\bigskip
\centerline {\smc 4. confrontation of predictions with observations}
\medskip

To assess the applicability of the binary broad-line region model to the
double-peaked Balmer lines of the target objects, we attempted to fit the radial
velocity curves constructed above to the equations of the double-lined
spectroscopic binary. In the convention adopted here, these equations can be
written as
\newcount\QutA \advance\Q by 1 \QutA=\Q
\newcount\QutB \advance\Q by 1 \QutB=\Q
$$
\eqalignno{
u_1 (t) = &\; (v_1\; \sin i)\; \sin \left[{2\pi\over P} (t-t_0) \right]
& (\the\QutA) \cr
u_2 (t) = &\; -(v_2\; \sin i)\; \sin \left[{2\pi\over P} (t-t_0)
\right]\; , & (\the\QutB) \cr
}
$$
where $t_0$ is the time at which the observed velocity of the two
peaks vanishes (inferior conjunction of the primary).  In the
cases where radial velocity curves for both peaks are available, these
were fitted simultaneously assuming a common period and phase and
allowing the two velocity amplitudes to be independent. This method,
in which all of the data are used simultaneously, produces the most
sensitive constraints on the model parameters. Because the period $P$
and zero phase epoch $t_0$ are nonlinear parameters, the fitting
algorithm scans the $P-t_0$ plane systematically. At each point the
values of the velocity amplitudes ($v_1\sin~i,~v_2\sin~i$) are
computed analytically by $\chi^2$ minimization, along with the value
of $\chi^2$ itself. The $P-t_0$ plane is explored thoroughly in order
to find the asymptotic behavior of the model parameters at very long
periods.

The resulting $\chi^2$ surface is used to find the best fit through
the velocity curves as well as to set limits on the model parameters
(see Lampton, Margon, \& Bowyer 1976). In particular, we determine the
shortest period that is consistent with the data (at 99\%\ confidence)
along with the corresponding velocity amplitudes. The velocity
amplitudes are linear parameters in the model adopted here; hence
their optimal values and confidence limits can be computed
analytically at every point in the $P-t_0$ grid. These are essential
ingredients for constraining the mass of the binary according to
inequalities (\the\QmlimA) and (\the\QmlimB). The mass ratio is also
determined; it is the ratio of the best-fitting radial velocity
amplitudes.

The algorithm described above was used to fit the annually-averaged
radial velocity curve for each object (see \S3). In Figure~\the\Ffit\
we show these velocity curves, with the best
fitting binary orbit models superposed. As this figure shows, the
best-fitting orbital motion models provide very poor descriptions of
the data. The reduced $\chi^2$ values corresponding to these fits are
considerably larger than unity (see Table~1), and thus the models are
formally rejected with very high confidence (evem with the generous
error bars used). Moreover, in the case of
3C~390.3 and 3C~332, the variability does not even approximate the
behavior of a spectroscopic binary. The two peaks do {\it not} drift
in opposite directions, and as a result the instantaneous velocity
ratio fluctuates considerably. If we insist on following the binary
black hole hypothesis to its logical conclusion, we would have to
allow for the existence of an additional source of noise superimposed on
the radial velocity curve that is larger than our original error bars.
We have investigated whether we can compensate for this hypothetical
shortcoming of our error analysis by renormalizing the reduced
$\chi^2$ surface to unit minimum.  This is equivalent to enlarging all
error bars until the radial velocity model of equations (\the\QutA)
and (\the\QutB) becomes consistent with the data. But then the
99\%-confidence lower limit on the orbital period from the
renormalized $\chi^2$ surface produces a fit that bears even less
resemblance to the original data, essentially because any evidence for
binary motion is lost in the much larger noise of unknown origin.  For
this reason, we have little confidence in the validity of this
renormalization procedure, and we do not adopt it the remaining
analysis.
\topinsert
\centerline{\smc TABLE 1: Lower Limits on Binary Orbit Parameters\ft{a}}
\medskip
\centerline{\vbox{\halign{
# \hfil \tabskip 1em & 
\hfil # \hfil \tabskip 1em & \hfil # \hfil \tabskip 1em & \hfil # \hfil \tabskip 1em \cr
\noalign{\hrule \vskip 2pt \hrule \vskip 1em}
 & Arp 102B & 3C 390.3 \hfil & 3C 332 \cr
\noalign{\vskip 1em \hrule \vskip 1em}
$\chi^2_\nu$ (d.o.f.)           & 8.54 (10)    & 18.5 (20)    & 0.74 (9) \cr
Mass Ratio                      & 1.5\ft{b}    & 1.20         & 1.33 \cr
Best-Fitting Period (yr)        & 390          & $\infty$     & 144 \cr
Minimum Period (yr)             & 159          & 811          & 82 \cr
Minimum Total Mass  (\Msol)     & 1.3\tten{10} & 4.2\tten{10} & 1.5\tten{10} \cr 
Minimum Orbital Separation (pc) & 0.34         & 1.47          & 0.23 \cr
\noalign{\vskip 1em \hrule \vskip 2pt \hrule \vskip 1em}
}}}
\centerline{\vbox{\hsize=5.5truein 
\item{\ft{a}} Minimum orbital periods are
derived by fitting the annually averaged velocity
curves, and correspond to the 99\% confidence levels in the appropriate
$\chi^2$ surface. The minimum mass and minimum orbital separation are 
then derived from the minimum period and the fitted velocity amplitude.
The most conservative (lowest) limits are listed here. 
\item{\ft{b}} The mass ratio for Arp~102B cannot be measured precisely
because its red peak is contaminated by the [S\ii] lines.  We
estimate $q \approx 1.5$, which we use to derive
the limits quoted here.
}}
\medskip
\endinsert

\pageinsert
\centerline{
\hbox to 6.5truein {\vsize=3.5truein \hskip -0.5truein
\psfig{figure=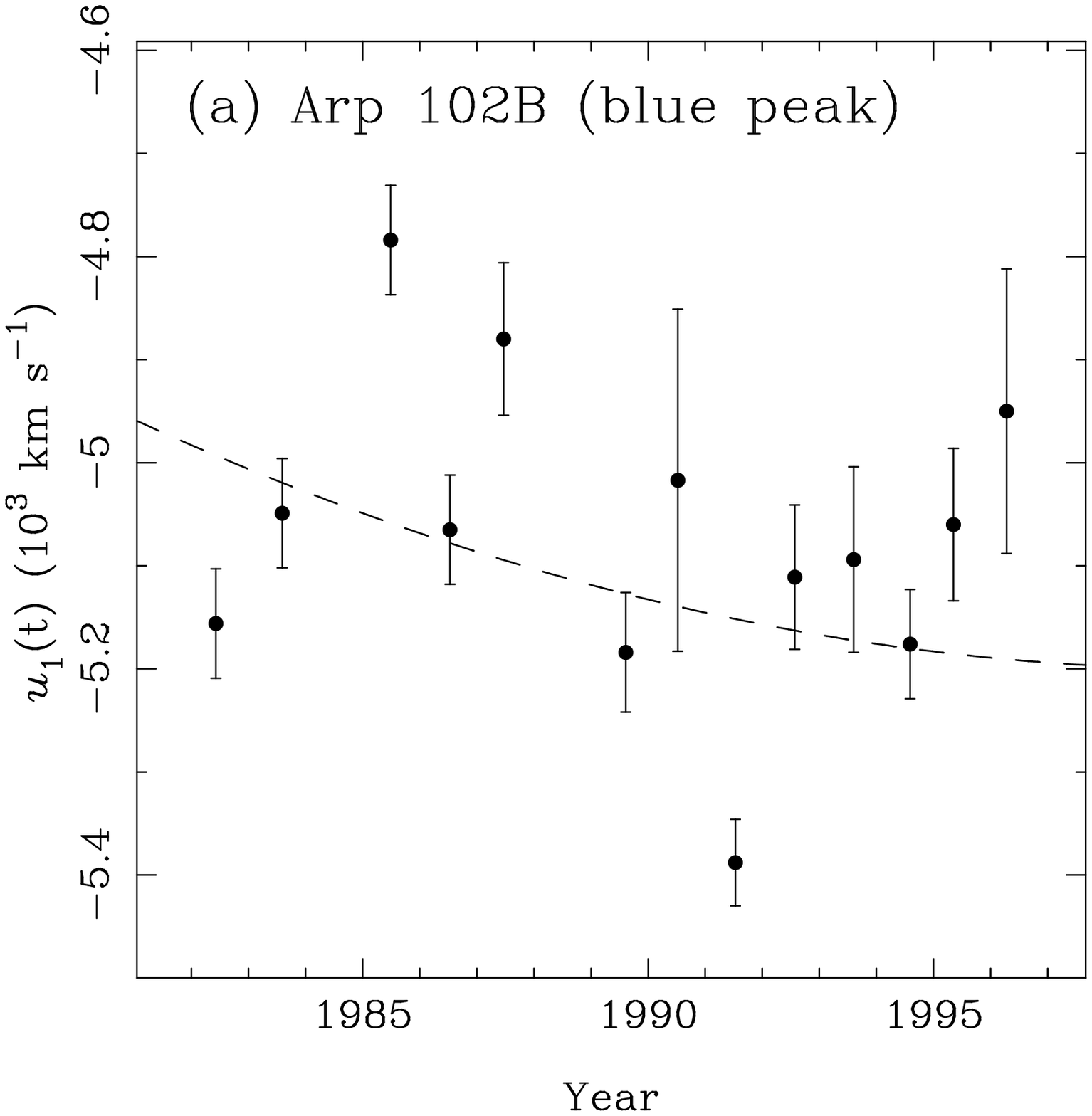,height=4.5in,rheight=3.5in,rwidth=3.4in}
\vbox to 3 truein {\hsize=3.5 truein
\noindent {\smc Figure \the\Ffit -- }
Annually-averaged radial velocity curves with the best-fitting binary models
(dashed lines) overlaid. The data points represent the avearge of all 
mesurements made from both the H\a\ {\it and} the H\b\ profiles from the 
same year. (a) In the case of Arp~102B, only the fit to the radial
velocity curve of the blue peak is shown since these are the only reliable data.
In the case of the other two objects, the simultaneous fits to the radial
velocity curves of both peaks are shown. The lower panels in (b) and (c) show
the variation of the red/blue velocity ratios with time; the horizontal dashed
line shows the velocity ratio derived from the best-fitting model. The error
bars in (a) and (b) are computed statistically as described in the text; they
represent the dispersion within a year rather than the  error in the annual
mean. The error bars in (c) represent the uncertainties in individual
measurements.
}\hfill}}
\centerline{
\hbox to 6.5truein {\hskip -0.75truein
\psfig{figure=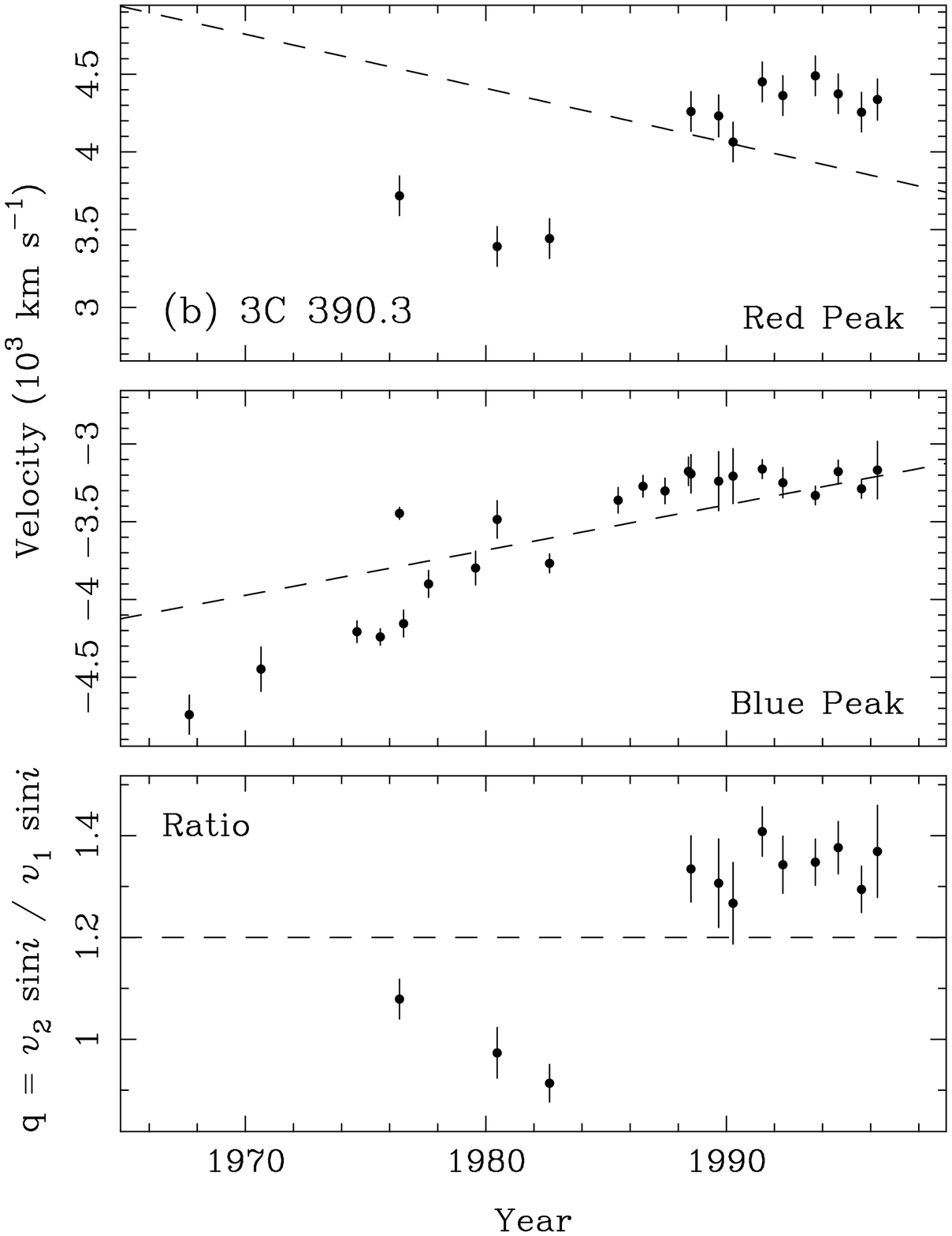,height=5.in,rheight=4.3in,rwidth=3.3in}
\psfig{figure=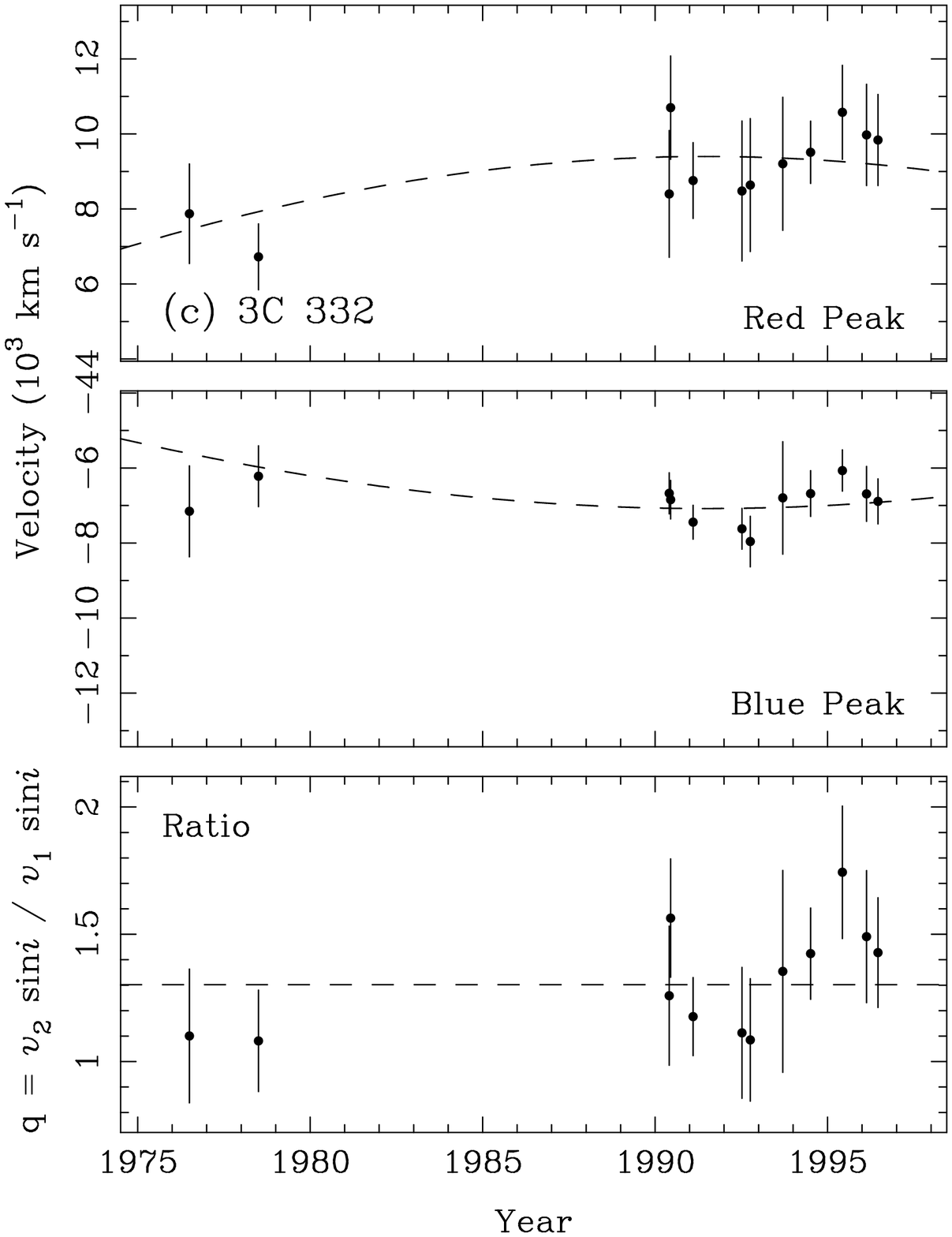,height=5.in,rheight=4.3in,rwidth=3.3in}
\hfill
}} \vfill
\endinsert

Regardless of whether or not an observed process is truly periodic,
there is an obvious limitation to any method that attempts to fit a
periodic function to data that span only a small fraction of the
hypothesized period. Any noise superposed on an otherwise smooth
radial velocity curve, such as that which is evident in Arp~102B and
apparently in 3C 390.3 as well, could grossly change the derived value
or lower limit to the period. This is a weakness of both Gaskell's
study of 3C~390.3 and ours. The lower limits on the period and the
mass that we derive are not necessarily secure ones if there are,
superimposed on the binary motion, occasional velocity variations of
unknown origin with a typical amplitude of $10^3$~km~s$^{-1}$. The
possibility of such contamination constitutes a caveat to our
otherwise clean rejection of the binary broad-line region model for
the three objects. However, because we are unable to suggest a secondary
mechanism for such large radial velocity variations, we find it more
reasonable to conclude that the observed changes are due to a single
as yet unexplained cause, for which binary orbital motion is an
unlikely explanation.

The lower limits to the orbital periods and the corresponding
projected velocity amplitudes were used to determine the lower limit
on the mass in each case. These limits are summarized in Table~1.
The corresponding mass ratios and lower
limits on the binary separation are also given. In the following
paragraphs we discuss the results for each object individually, paying
attention to its idiosyncrasies.

\medskip
\centerline{\it 4.1. Arp 102B}
\medskip
The minimum total mass obtained from fitting the velocity curve of
Arp~102B is $5.1 \times 10^9\,\Msol$, under the assumption that
$q=1$. This is higher than that reported by Halpern \& Filippenko
(1988, 1992), which is an expected benefit of the lengthening time
span of the observations. The constraints that we derive for Arp~102B
are not as restrictive as they might have been, because we still do
not make use of the red peak of its H\a\ line.  The velocity of the
red peak cannot be measured reliably because it coincides with the
[S\ii] \tl6717, 6731 doublet.  This complication also prevents us from
assigning a precise value to the mass ratio. Nevertheless, we can
conclude from the asymmetry of the line profile and from the apparent
coincidence of the red peak of H\a\ with the [S\ii] doublet that the
hypothesized mass ratio must be about 1.5, similar to the two other
objects whose red peaks are easily measured. If so, the minimum mass
would be 1.3\tten{10}~\Msol, about twice as large as that derived
under the conservative assumption of equal black hole masses. In
Table~1 we report the limits that correspond to $q=1.5$, which we
consider a more reasonable and realistic assumption than $q=1$.

The radial velocity curve of Arp~102B presented in Figure~\the\Frvel$a$ shows
considerable scatter, of order 300~km~s$^{-1}$ about a mean of 5100~km~s$^{-1}$. 
As we have argued, this is likely the result of intrinsic profile variations on
short time scales which probably dominate the uncertainty in the location of
the peak. If we use this argument as a basis for rejecting some data points from
the analysis, a minimum mass can be derived that is higher than those we report
in Table~1.  This is a second reason why the lower mass limits listed for
Arp~102B in Table~1, which are derived using {\it all} of the points in each
velocity curve, are the most conservative (lowest) that one could infer.  In
summary, we consider $10^{10}\,\Msol$ a realistic lower limit on the total
mass of the black holes in Arp~102B in the binary broad-line region scenario.

\medskip
\centerline{\it 4.2. 3C 390.3}
\medskip
The radial velocity curve of the {\it blue} H\b\ peak of 3C~390.3 presented by
Gaskell (1996$a$) shows that the peak drifted systematically by about 1700~\kms\
between 1968 and 1988. This behavior was interpreted by Gaskell as the signature
of orbital motion, with a most likely period of 300 years and a total binary
mass of $6.6 \times 10^9\,\Msol$. Our newer data, however, show that this trend
did not continue after 1988. To illustrate the subsequent deviation, we plot in
Figure~\the\FGask\ the data and best binary orbit model of Gaskell (1996$a$) and
superpose our additional data for comparison. It is clear from this figure that
the velocities after 1988 do not follow the extrapolation of the binary model
that fits the velocities between 1968 and 1988. The direction of the deviation
disfavors {\it any} orbital model for the radial velocity variations observed
over the full time span. Furthermore, the observed trend in the velocity of the
{\it red} peak of H\a\ is inconsistent with orbital models: in the binary
orbital motion picture, the two peaks should drift in opposite directions, which
is flatly contradicted by the observations (Figure~\the\Ffit$b$).
\topinsert
\vskip -0.6truein
\centerline{
\psfig{figure=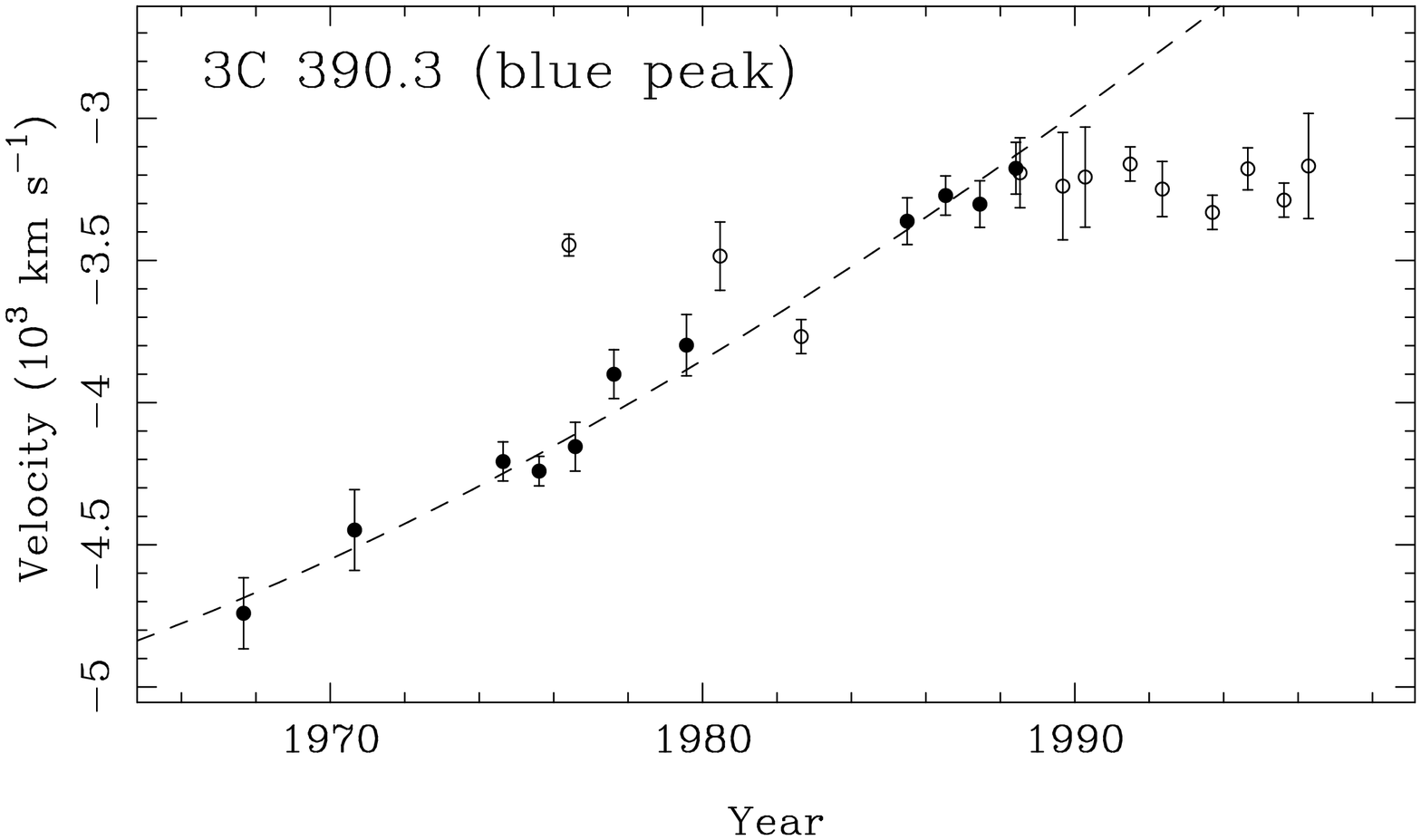,width=5in,rheight=3.3in,angle=-90}
}
\centerline{\vbox{\hsize=5.5truein \noindent{\smc Figure~\the\FGask\ -- }
An expanded view of the radial velocity curve of the blue peak of
3C~390.3. The filled circles are the data of Gaskell (1996$a$) with
error-in-the-mean error bars, while the open circles are our own data
with r.m.s.-dispersion error bars.  The dashed line shows Gaskell's
fit to the data prior to 1988; it corresponds to a period of
300~yr. The radial velocities measured after 1988 clearly do not
follow the extrapolation of the fit based on data prior to 1988.
}}
\medskip
\endinsert

The year 1989 coincided with a revealing development in the line profile of
3C~390.3 which constitutes another obstacle to the binary broad-line region
interpretation. Previously the displaced peaks had round tops, and were easily
located by fitting symmetric functions. But in 1989 the blue peak developed the
sharp cusp shown in Figure~\the\Fpec . We are confident in the reality of this
feature because it lies 37~\AA\ blueward of the head of the telluric
B-band, and because
it persisted through 1990.  By 1991, the blue peak had resumed its more normal,
rounded appearance.  In addition to complicating the measurement of ``a radial
velocity,'' this cusplike feature is evidence of a well ordered velocity field,
and not the random distribution of cloud orbits that is implicit in the binary
scenario. It is perhaps significant that the appearance of this transient
feature coincided with the ending of the radial velocity trend noticed by 
Gaskell (1996$a$).  Apparently, there exists a dynamical process, one not
consistent with binary orbital motion, that is responsible for shaping the line
profile.

Despite these warnings that a binary broad-line region model is unsuited to
3C~390.3, if we persist and fit its velocity curve to a binary orbit, we
obtain the period and mass limits given in Table~1.  The mass limit 
of 4.2\tten{10}~\Msol\ can be made
more restrictive by using information on the inclination of the orbital plane.
From the observations of superluminal motion in 3C~390.3 by  Alef et al.
(1994, 1996), Eracleous, Halpern, \& Livio (1996) derive an upper limit to 
the inclination angle of the jet to the line of sight of $i < 33$\deg.
Assuming that the plane of the orbit is perpendicular to the axis of the radio
jet, this limit implies that $M > 2.6 \times 10^{11}$~\Msol, a value 40
times larger than that derived by Gaskell\ft{5} and large enough to render the
assumed model for the emission line implausible (see \S5).
\footnote{}{\parindent=0pt \item{\ft{5}}
The derivation of the mass of the binary by Gaskell (1996$a$) also took into
account the inclination of the orbit using a jet inclination of $i = 29$\deg.
That estimate was made by Ghisellini et al. (1993) from a combination of
superluminal motion data and a synchrotron self-Compton model for the X-ray
emission. The method is similar but not identical to that of Eracleous et al.
(1996), and the data were taken from an independent source. The resulting
estimates are consistent with each other.}

\medskip
\centerline{\it 4.3. 3C 332}
\medskip

Because 3C~332 was not observed as frequently as the other two targets, it is
not possible to compute statistical error bars (using the annual dispersion of
individual measurements) for all of its radial velocity data points. We
therefore adopt the individual measurement uncertainties. Figure~\the\Ffit$c$
shows these measurements and associated error bars, along with the best-fitting
binary orbit model. This is the only object for which a binary orbit model
produces a formally acceptable fit, mainly because of the large error bar
associated with each measurement. Even though the velocity curve of 3C~332 is
not as well sampled as those of the other two objects, it still yields an
interesting mass constraint, $M > 1.5 \times 10^{10}$~\Msol, on the total mass
of a supermassive binary that this object may harbor. Two observational factors
combine to achieve this sensitive limit. First, the time spanned by the
radial velocity curve is quite long, and second, the separation between the
twin peaks of H\a\ is almost twice that of the other two objects. We note 
that the large velocity separation between the two peaks is {\it not} 
necessarily an  indication of an extremely massive binary or a small
binary  separation. The {\it observed} velocity separation between the two 
peaks is 
\newcount\Qvsep \advance\Q by 1 \Qvsep=\Q
$$
\Delta v_{\rm peaks} = \left({G M \over a}\right)^{1/2}\; 
\sin i \;\; \sin \varphi ,
\eqno{(\the\Qvsep)}
$$
where $a$ is the binary separation and $\varphi$ is the orbital phase 
($\varphi=0$ corresponds to inferior
conjunction of the primary). It is clear from equation (\the\Qvsep) that  the
velocity widths of the lines depend at least as sensitively on the observer's
perspective (i.e., the orbital inclination and phase) as they do 
on the intrinsic properties of the binary (mass and separation).

\bigskip
\centerline {\smc 5. discussion}
\medskip
A binary broad-line region interpretation of double-peaked emission lines
implies, through the simple dynamical arguments of the preceding sections,
binary black hole masses greater than $10^{10}\,\Msol$ in all three of the
objects studied. The mass limits that we derive are the lowest 
reasonable limits, since throughout our analysis we have tried to err on the
side of caution. These very large required masses render a binary
interpretation very unlikely for a number of reasons. In this section, we
discuss how these assumed masses conflict with other observations and with
theory. 

First, a mass of the order of \ten{10}~\Msol\ is well in excess of the
masses of central black holes in active and non-active galaxies that have been
measured by means of circular gas velocities or stellar kinematics. Examples of
such mass determinations in {\it active} galaxies include 3.5\tten{7}~\Msol\
in NGC~4258 (Miyoshi et al. 1995), 5\tten{8}~\Msol\ in NGC~4261 (Ferrarese,
Ford, \& Jaffe 1996), \ten{9}~\Msol\ in NGC~4594 (Kormendy et al. 1996$a$),
and 2\tten{9}~\Msol\ in M87 (Harms et al. 1994). Examples of  non-active
galaxies with a well-determined nuclear black hole mass are NGC~3115 ($M\approx
2\times 10^9$~\Msol; Kormendy et al. 1996$b$), NGC~4486B ($M\approx 6\times
10^8$~\Msol; Kormendy et al. 1997), and M32 ($M\approx 3\times
10^6$~\Msol; van der Marel et al. 1997). Certainly these observational
techniques suffer no technical difficulties that would prevent them from
detecting single black holes of \ten{10}~\Msol. In fact, one would expect 
very massive black holes to be easily detectable by the above techniques. Why 
should the most massive black holes be found only in binaries?

A second objection to such extremely massive binaries is the remarkable coincidence
that would be required for {\it two} record-setting black holes, of nearly equal
mass, to find themselves in the same galaxy and emitting nearly the same
Balmer-line luminosity.  One such system could be dismissed as unique and
unrepresentative, but there are several such systems in our neighborhood.  It is
more reasonable to interpret the approximate symmetry in velocity and flux of
these double-peaked emission lines as the workings of a {\it single}, less
massive object, even though we may not yet understand the process that
is responsible.

A third, and potentially serious weakness of the binary model is that the very
large requisite masses strain against {\it upper} limits that can be derived
from variability time scales, most notably via X-ray observations. If we accept
that one or both of the members of the hypothesized binary is responsible for
the AGN's X-ray emission and its variability, then the size $r$ of either
emission region can be no greater than $c\Delta t$, where $\Delta t$ is the
observed variability time scale.  Because the X-ray emitting region is thought
to occupy the inner parts of an accretion disk, up to $r\approx 10\; r_{\rm g}$
(where $r_{\rm g}\equiv GM/c^2$), the implied upper limit to the mass
of the black hole is
\newcount\Qlight \advance\Q by 1 \Qlight=\Q
$$
M < 1.7\times 10^9 \; \left({\Delta t \over {\rm 1~day}}\right)\; 
\left({r\over 10\,r_{\rm g}}\right)^{-1} \Msol.
\eqno{(\the\Qlight)}
$$
One of the best studied AGN X-ray light curves is that of
3C~390.3. During long-term monitoring by {\it ROSAT}, 3C 390.3
displayed large-amplitude flares with a minimum doubling time scale of
9 days (Leighly et al. 1997$a,b$).  This behavior implies a maximum
black hole mass of $1.5 \times 10^{10}\,\Msol$, uncomfortably low
compared to the lower limit in Table~1.  One can relieve this strain
by invoking relativistic motion for the source of the X-rays, which
would make the {\it intrinsic} variability time scale longer than the
observed one. However, the similarity between the X-ray spectra of 3C
390.3 and those of ordinary Seyfert galaxies, most notably the
presence of an Fe~K$\alpha$ line at the rest energy in the frame of
the galaxy (Eracleous et al. 1996), argues that the X-ray source in 3C
390.3 is not beamed.  We consider the observed rapid X-ray variability
to be a strong argument against the association of the X-ray source
with a black hole more massive than $10^{10}\,\Msol$ in 3C~390.3, and
certainly against masses greater than $10^{11}\,\Msol$ that would be
required if the orbital inclination were less than $33^{\circ}$. The
X-ray variability properties of our other two objects have not yet
been studied, but similar monitoring observations of Arp~102B and
3C~332 could be very effective in disallowing the large masses
required for them by the binary broad-line region model.

A fourth issue that needs to be revisited if the mass of the hypothesized binary
is of order \ten{10}~\Msol\ or larger is the decay of the orbit by
gravitational radiation, which is the final process in the coalescence of the
binary.  The expression for the time to coalescence given by Lightman et
al. (1979) can be written as
\newcount\Qlife \advance\Q by 1 \Qlife=\Q
$$
T_{\rm gr} = 5.7\times 10^8\;\; {(1+q)^2\over q}\;\;
{a_{\rm pc}^4 \over M_{10}^3}
\;\; {\rm yr},
\eqno{(\the\Qlife)}
$$
where $M_{10}$ is the mass of the binary in units of \ten{10}~\Msol\ and
$a_{\rm pc}$ is the binary separation in pc. Upper limits on the binary
coalescence times can be obtained using the minimum masses derived here and
upper limits on the binary separation from VLBI observations of the radio cores
of these objects. In particular, observations of 3C~390.3 and Arp~102B by Alef
et al. (1994, 1996) and Biermann et al. (1981) show that the radio cores are
unresolved and yield upper limits to the core sizes of 1.4~pc and 0.4~pc
respectively. The corresponding VLBI maps have very high dynamic range so that
close radio doubles with a large intensity contrast would have been detected had
they been separated by more than the core sizes given above.  The implied upper
limits to the coalescence times are 7\tten{7}~yr for Arp~102B and 3\tten{8}~yr
for 3C~390.3.  If instead the orbital separation of Arp~102B is as small as
that given in Table~1, which is highly desirable to minimize the mass, the
coalescence times could be as small as 9\tten{6}~yr. The corresponding
coalescence time of 3C~332 is only 2\tten{6}~yr. Since the coalescence time is
such a sensitive function of the mass of the system, it is surprising that 
all three objects in our collection have large masses. Objects with masses 
that are even a factor of 2 smaller would have almost an order of magnitude
longer coalescence times and hence they would be more likely to be found.
We note that the observed velocity separation of the two peaks does
not necessarily imply a shorter lifetime [cf., equation (\the\Qvsep)].
Moreover, in the case of 3C~390.3 the inferred minimum separation of the 
hypothesized binary is comparable to the spatial resolution of the available
VLBI map, which implies that the binary should have been resolved. If the 
mass of the binary in 3C~390.3 is indeed 2.6\tten{11}~\Msol, according to
the upper limit on the inclination, the binary separation is 2.3~pc,
easily resolvable in the VLBI map of Alef et al. (1994, 1996). 

There are additional problems with the binary broad-line region interpretation
that are unrelated to the large masses required. One that has been recognized
for some time was first noted by M. V. Penston, and summarized in footnote 3 of
Chen, Halpern, \& Filippenko (1989). The line profile of a binary broad-line
region should not consist of two widely displaced peaks, because the velocity
dispersion of gas bound to each black hole should be {\it greater} than its
binary orbital velocity.  Thus, the double line profile should be more highly
blended.  In defense of the model, the possibility of fine tuning between the
opposing forces of gravity and radiation pressure has been invoked to decrease
the line width of each individual broad-line region (Gaskell 1996$a$, drawing
from the model of Mathews 1993). We are unable to judge how reasonable it is to
assume such a balance.

Finally, recent results from {\it HST\/} ultraviolet spectroscopy of Arp~102B
provide a new and striking view of this object that offers no support for a
binary broad-line region interpretation (Halpern et al. 1996). In all the
ultraviolet (and visible) emission lines, there is present an ``ordinary''
broad-line component at zero velocity.  In contrast, the double-peaked component
is strong only in the Balmer lines and in Mg\ii, but is completely absent in
Ly~$\alpha$ and other high-ionization lines, indicating an unusually low level
of ionization in the double-peaked components. Evidently, extreme physical
conditions are present in the high velocity gas.  A model in which two black
holes are surrounded by individual broad-line regions does not account for the
unusual line intensity ratios in either of the displaced peaks of Arp~102B.

\bigskip
\centerline {\smc 6. epilogue}
\medskip
Although we have found no evidence to support the binary broad-line region
interpretation of double-peaked emitters from long-term monitoring of three
objects, this does not mean that a binary model cannot account for the emission
line profiles of some of the many other AGNs with either single or double
displaced peaks.  Our three targets were chosen precisely because the very large
velocity widths of their lines enable meaningful results to be obtained in just
a couple of decades.  Emission-line peaks with velocities significantly less
than these, such as in OX 169 (Stockton \& Farnham 1991), may still arise in a
less massive spectroscopic binary, even if the peaks appear stationary for many
years. OX~169 and similar objects remain plausible binary candidates.  We do 
conclude, however, that the {\it transient} double-peaked emission lines
that were recently discovered in an additional trio of objects, namely NGC~1097
(Storchi-Bergmann, Baldwin, \& Wilson 1993; Storchi-Bergmann et al. 1995, 1997),
Pictor~A (Halpern \& Eracleous 1994; Sulentic et al. 1995), and M81 (Bower et
al.  1996), are {\it not} likely to originate in binary broad-line regions. 
These suffer most of the same objections that apply to the persistent
double-peaked emitters. Their peaks are too symmetric and widely spaced, and at
least two of these objects, M81 and Pictor~A, have long possessed ordinary broad
lines at zero velocity, as well as luminous and variable X-ray emission that
does not appear to have increased in concert with their new emission-line
components. In addition, the very transient nature of these displaced peaks,
appearing and disappearing together as they do, argues against an origin in
independent accreting systems.  Because these transient double-peaked emitters
are so nearby, it seems likely that they are quite common among AGNs,
outnumbering the persistent ones by a large factor. Our understanding of both
groups would greatly benefit if they could be ``unified'' by the same mechanism.

We emphasize that even in the three objects for which we have derived strong
constraints against the binary broad-line region model, we have not ruled out
the {\it existence} of a supermassive binary black hole, but only the scenario
in which separate black holes are responsible for the two displaced
emission-line peaks.  It is still possible that binaries {\it less} massive than
the limits listed in Table~1 lurk in these AGNs, if the emission-line source is
associated with only one of the black holes. The observational signature of such
a system would be an equal modulation of the velocities of the two peaks in the
{\it same} direction. We have not yet seen any such evidence of a {\it
single-lined} spectroscopic binary, although the relative motions of the two
emission-line peaks of 3C~390.3 shown in Figure~\the\Ffit$b$ might be compatible
with such a description.  Taking into account possible random velocity
variations of $K \simeq 1000$~\kms\ resulting from unexplained changes in the
line profiles, we estimate that we would be sensitive to systems with mass
ratios $q<12\; (K/1000~{\rm km~s^{-1}})^{-1}\; (M_9/a_{17})^{1/2}$ for orbital
periods comparable to the time span of our data. Another possibility which 
remains open
is that some AGNs harbor supermassive binaries with a single, circumbinary
broad-line region. We speculate that an observational signature of these systems
would be
periodic episodes of brightening caused by a modulated accretion rate (c.f.,
Artimowicz \& Lubow 1996). 

Even though we judge that the arguments against the binary broad-line
region model for the objects that we have studied are already quite
strong, there are powerful additional tests that could be
performed. First, there exists a large amount of spectroscopic data on
3C~390.3 obtained in conjunction with a one-year reverberation
monitoring campaign (Dietrich et al. 1997).  If the H$\alpha$ emission
line shows reverberation in response to continuum variations, and if
{\it both} sides of the line respond with little or no delay between
them, then a binary broad-line region model is virtually ruled out,
since the required orbital separation is 0.3~pc in Gaskell's (1996$a$)
analysis, or greater then 1.5~pc in ours.  We eagerly await the
results of these observations. Second, it is possible in principle to
detect absolute motions of compact cores with a precision of tens of
microarcseconds using relative astrometry between widely spaced VLBI
sources (Lara et al. 1996).  If the compact core of an object like
3C~390.3 were really moving at a velocity of 5,000~km~s$^{-1}$
relative to its host galaxy, then a displacement of 0.1
milliarcseconds would be expected in 30 years.  In a time comparable
to the length of our optical spectroscopic studies, independent
evidence for or against binary orbital motion could thus be obtained
from VLBI astrometry.

We have refrained in this paper from speculating upon the actual
origin of these double-peaked emission lines.  That is a much more
difficult task than the falsification of one particular, albeit
important, model. A discussion of more plausible models is presented
in those same papers that contain the detailed descriptions of our
spectra and their variability (Newman et al. 1997; Gilbert et
al. 1997).

\bigskip
\centerline {\smc acknowledgments}
\medskip \noindent
We are indebted to all the Lick Observatory observers who allowed us
to use old data that they have been saving for the past two
decades. In particular, we are grateful to D. E. Osterbrock,
M. M. Phillips, and S. A. Grandi.  We also thank W.  Zheng for passing
on many of the old Lick spectra to us in a convenient, electronic
form, and the referee, C. M. Gaskell, for his thoughtful comments.
M.E. acknowledges support by Hubble Fellowship grant
HF-01068.01-94A. A.M.G. and J.A.N. acknowledge support from a Eugene
Cota-Robles and a National Science Foundation Fellowship
respectively. This work was also partly supported by NASA grant
G0-06097-94A from the Space Telescope Science Institute (operated by
AURA, Inc., under NASA contract NAS5-26555).

\goodbreak
\bigskip
\centerline {\smc references}
\medskip

\ref {Alef, W., Preuss, E., Kellerman, K. I., Wu, S. Y., \& Qiu, Y. H. 1994,
      in Compact Extragalactic Radio Sources, NRAO Workshop No. 23, ed.
      J. A. Zensus \& K. I. Kellerman (Green Bank: NRAO), 55}
\ref {Alef, W., Wu, S. Y., Preuss, E., Kellerman, K.~I., \& Qiu, Y.~H. 1996,
      A\&A, 308, 376}
\ref {Artimowicz, P. \& Lubow, S. H. 1996, ApJ, 467, L77} 
\ref {Begelman, M. C., Blandford, R. D., \& Rees, M. J. 1980, Nature, 287,
      307}
\ref {Biermann, P., Preuss, E., Kronberg, P. P., Schilizzi, R. T., \&
      Shaffer, D. B. 1981, ApJ, 250, L49}
\ref {Bower, G. A., Wilson, A. S., Heckman, T. M., \& Richstone, D. O. 1996,
      AJ, 111, 1901}
\ref {Chen, K., Halpern, J. P., \& Filippenko, A. V. 1989, ApJ, 339, 742}
\ref {Dietrich, M. et al. 1997, ApJ, submitted}
\ref {Eracleous, M. 1997, Adv Space Res, in press}
\ref {Eracleous, M., \& Halpern, J. P. 1994, ApJS, 90, 1}
\ref {Eracleous, M., Halpern, J. P., \& Livio, M. 1996, ApJ, 459, 89}
\ref {Ferrarese, L., Ford, H. C., \& Jaffe, W. 1996, ApJ, 470, 444}
\ref {Gaskell, C. M. 1983, in Proc. 24th Li\`ege Int. Astrophys. Colloq.,
      Quasars and Gravitational Lenses (Cointe-Ougree: Univ. Li\`ege), 473}
\ref {\ditto\ . 1988, in Active Galactic Nuclei, ed. H. R. Miller \&
      P. J. Witta (Berlin: Springer), 61}
\ref {\ditto\ . 1996$a$, ApJ, 464, L107}
\ref {\ditto\ . 1996$b$, in Jets from Stars and Galactic Nuclei, ed. W. Kundt,
      (Berlin: Springer), 165}
\ref {Gaskell, C. M. \& Snedden, S. A. 1997, in Emission Lines in Active 
      Galaxies: New Methods and Techniques, ed. B. M. Peterson, F.-Z. Cheng, 
      \& A. S. Wilson (San Francisco: ASP), 193}
\ref {Ghisellini, G., Padovanio, P., Celotti, A., \& Maraschi, L. 1993, 
      ApJ, 407, 65}
\ref {Gilbert, A. M., Eracleous, M., Halpern, J. P., \& Filippenko, A. V. 1997,
      in preparation}
\ref {Grandi, S. A., \& Osterbrock, D. E. 1978, ApJ, 220, 783}
\ref {Grandi, S. A., \& Phillips, M. M. 1979, ApJ, 232, 659}
\ref {Halpern, J. P., \& Eracleous, M. 1994, ApJ, 433, L17}
\ref {Halpern, J. P., Eracleous, M., Filippenko, A. V., \& Chen, K. 1996,
      ApJ, 464, 704}
\ref {Halpern, J. P., \& Filippenko, A. V. 1988, Nature, 331, 46}
\ref {\ditto\ . 1992, in Testing the AGN Paradigm,
      ed. S. S. Holt, S. G. Neff, \& C. M. Urry, AIP Conf. Proc. 254, 
      (New York: AIP), 57}
\ref {Harms, R., et al., 1994, ApJ, 435, L35}
\ref {Hoffmeister, C., Richter, G., \& Wenzel, W. 1985, Variable Stars (Berlin:
      Springer)}
\ref {Kormendy, J., et al. 1996$a$, ApJ, 459, L57}
\ref {\ditto\ . 1996$b$, ApJ, 473, L91}
\ref {\ditto\ . 1997, ApJ, 482, L139}
\ref {Lampton, M., Margon, B., \& Bowyer, S. 1976, ApJ, 208, 177}
\ref {Lara, L., Marcaide, J. M., Alberdi, A., \& Guirado, J. C. 1996, A\&A,    }
\ref {Lauer, T. R., et al. 1991, ApJ, 369, L41}
\ref {\ditto\ . 1992, AJ, 103, 703}
\ref {\ditto\ . 1995, AJ, 110, 2622}
\ref {Leighly, K. M., \& O'Brien, P. T. 1997$a$, ApJ, 481, L15}
\ref {Leighly, K. M., O'Brien, P. T., Edelson, R., George, I. M., Malkan, M. A.,
      Matsuoka, M., Mushotzky, R. F., \& Peterson, B. M. 1997$b$, ApJ, 483, 767}
\ref {Lightman, A. P., Press, W. H., Price, R. H., \& Teukolsky, S. A. 1979,
      Problem Book in Relativity and Gravitation, (Princeton: Princeton U.
      Press), 495}
\ref {Mathews, W. G. 1993, ApJ, 412, L17}
\ref {Miyoshi, M., Moran, J. M., Herrnstein, J. R., Greenhill, L. J.,
      Nakai, N., Diamond, O. J., \& Inoue, M. 1995, Nature, 373, 127}
\ref {Newman, J. A., Eracleous, M., Halpern, J. P., \& Filippenko, A. V., 1997,
      ApJ, in press}
\ref {Oke, J. B. 1987, in Superluminal Radio Sources, ed. J. A. Zensus \& T. J.
      Pearson (Cambridge: Cambridge University Press), 267}
\ref {Polnarev, A. G., \& Rees, M. J. 1994, A\&A, 283, 301}
\ref {Quinlan, G. D. 1996, New Astronomy, 1, 35}
\ref {Stauffer, J., Schild, R., \& Keel, W. 1983, ApJ, 270, 465}
\ref {Stockton, A., \& Farnham, T. 1991, ApJ, 371, 525}
\ref {Storchi-Bergmann, T., Baldwin, J. A., \& Wilson, A. S. 1993, ApJ, 410, L11}
\ref {Storchi-Bergmann, T., Eracleous, M., Livio, M., Wilson, A. S.,
      Filippenko, A. V., \& Halpern, J. P.  1995, ApJ, 443, 617}
\ref {Storchi-Bergmann, T., Ruiz, T. R., \& Eracleous, M. 1997, Adv Space Res,
      in press}
\ref {Sulentic, J. W., Marziani, P., Zwitter, T., \& Calvani, M. 1995, ApJ,
      438, L1}
\ref {van der Marel, R. P., de Zeeuw, P. T., Rix, H.-W., \& Quinlan, G. D.
      1997, Nature, 385, 610}
\ref {Wade, R. A. \& Horne, K. 1988, ApJ, 324, 411}
\ref {Zheng, W., P\'erez, E., Grandi, S. A., \& Penston, M. V. 1995, AJ,
      109, 2355}

\end